\documentclass[twocolumn,showpacs,aps,epsfig,nofootinbib]{revtex4}
\usepackage{graphicx}
\usepackage{epstopdf}
\usepackage{latexsym}
\usepackage[center]{subfigure}

\begin{document}

 \newcommand{\Pvarphi}{{p_{\varphi}}}
 \newcommand{\rhovarphi}{{\rho_\varphi}}

%%%%%%%%%%%%%%%%%%%%%%%%%%%%%%%%%%%%%%%%%%%%%%%%%%%%%%%%%%%%%%%
 \newcommand{\bq}{\begin{equation}}
 \newcommand{\eq}{\end{equation}}
 \newcommand{\bqn}{\begin{eqnarray}}
 \newcommand{\eqn}{\end{eqnarray}}
 \newcommand{\nb}{\nonumber}
 \newcommand{\lb}{\label}
\newcommand{\PRL}{Phys. Rev. Lett.}
\newcommand{\PL}{Phys. Lett.}
\newcommand{\PR}{Phys. Rev.}
\newcommand{\CQG}{Class. Quantum Grav.}
 %%%%%%%%%%%%%%%%%%%%%%%%%%%%%%%%%%%%%%%%%%%%%%%%%%%%%%%%%%%%%%%

\title{Vector and tensor perturbations in Horava-Lifshitz cosmology}

\author{Anzhong Wang $^{b}$}
%\email{anzhong_wang@baylor.edu}

\affiliation{ GCAP-CASPER, Physics Department, Baylor University,
Waco, TX 76798-7316, USA  }

\date{\today}

\begin{abstract}

We study cosmological vector and tensor perturbations in Horava-Lifshitz gravity, adopting the most 
general  Sotiriou-Visser-Weinfurtner generalization without the detailed balance but with projectability 
condition. After deriving the general formulas in a flat FRW background, we find that the vector 
perturbations are identical to those given in general relativity. This is true also in the non-flat cases. 
For the tensor perturbations, high order derivatives of the curvatures produce effectively an anisotropic 
stress, which could have significant efforts on the high-frequency modes of gravitational waves, while
for the low-frenquency modes, the efforts are negligible. The power spectrum is scale-invariant in the 
UV regime, because of the particular dispersion relations. But, due to lower-order corrections, it will 
eventually reduce to that given in GR in the IR limit. Applying the general formulas to the de Sitter and 
power-law backgrounds, we calculate the power spectrum and index, using the uniform approximations, 
and obtain their analytical expressions in both cases.

\end{abstract}

\pacs{04.60.-m; 98.80.Cq; 98.80.-k; 98.80.Bp}

\maketitle

\section{Introduction}
\renewcommand{\theequation}{1.\arabic{equation}} \setcounter{equation}{0}

The background dynamics and the generation and evolution of
perturbations during a period of inflation in the early universe,
may deviate from the standard results if general relativity (GR)
acquires significant ultra-violet (UV) corrections from a quantum
gravity theory. Horava recently proposed such a theory
\cite{Horava}, motivated by the Lifshitz theory in solid state
physics \cite{Lifshitz}. Horava-Lifshitz (HL) theory has the
interesting feature that it is non-relativistic in the UV regime,
i.e., Lorentz invariance is broken. The effective speed of light
in the theory diverges in the UV regime, which could potentially
resolve the horizon problem without invoking inflation.
Furthermore, scale-invariant super-horizon curvature perturbations
could be produced without inflation \cite{Muk,KK,Piao,GWBR,YKN,KUY,WM,WWM}.

Originally, Horava assumed two conditions -- detailed balance and
projectability (though he also considered the case where the
detailed balance condition is softly broken) \cite{Horava}. 
%So far most of the work
%on the HL theory has abandoned the projectability condition but
%maintained detailed balance \cite{Cosmos,BHs,others}. One of the
%main reasons is that the resulting theory is much simpler to deal
%with, giving local rather than global energy constraints. However,
Later, it was found that breaking the projectability condition is problematic 
\cite{CNPS}
and gives rise to an inconsistent theory \cite{LP}. With detailed
balance, on the other hand, the scalar field is not UV stable
\cite{calcagni},
and the theory requires a non-zero negative cosmological constant
and breaks parity in the purely gravitational sector \cite{SVW}
(see also \cite{KK}). 

To resolve these problems, various modifications have been proposed 
\cite{BPS2,HMT}. In particular, 
the Sotiriou-Visser-Weinfurtner (SVW) 
generalization is the most general setup of the HL theory with the
projectability condition and without detailed balance \cite{SVW}.
The preferred time that breaks Lorentz invariance leads to a
reduced set of diffeomorphisms, and as a result, a spin-0 mode of
the graviton appears. This mode is potentially dangerous and may
be not stable, and cause strong coupling problems that could prevent the recovery of
GR in the IR limit \cite{CNPS,KA,PS}.
To address these issues and apply the theory to cosmology,
 % { linear} cosmological 
 linear perturbations of
the Friedmann-Robertson-Walker (FRW) model with arbitrary spatial
curvature in the SVW setup were studied,  and shown explicitly that the spin-0
scalar mode of the graviton is stable in both the IR and the UV
regimes, provided that $ 0\leq\xi\leq 2/3$, where $\xi$
is a dynamical coupling parameter  \cite{WM}. { However, This stability condition
has the unwanted consequence that the scalar mode is a ghost. %\cite{Horava,Mukb,BPS,KA}. 
To tackle this problem, one may
consider the theory in the range $\xi \le 0$, so that the ghost problem is avioded.
But, within this range, the spin-0 mode becomes unstable, since now the sound speed $c_s^2=\xi/(2-3\xi)$ is
non-positive. However in the limit that the sound speed becomes small, % as $\xi\to0$,
one should undertake a non-linear analysis to determine whether the
strong self-coupling of the scalar mode decouples, as in the Vainshtein mechanism
in massive gravity \cite{Vain}. Taking these non-linear effects into account, Mukohyama 
recently showed that the continuous limit, $\xi \rightarrow 0$, of GR indeed exists for spherically 
symmetric, static, vacuum configurations \cite{Mukc}. In this paper, we  assume that the strong-coupling 
 problem in the cosmological background can also be addressed via this mechanism  \cite{VainPRD} or 
 some other approach.

In addition, it could  be quite possible that the legitimate  background in the HL theory is 
not Minkowski. In particular,   recent observations show that our universe is currently de Sitter-like
\cite{obs}. Therefore, an alternative is to consider the de Sitter  space as the background. 
Along this direction of thinking, 
the stability of the de Sitter  spacetime in the SVW setup was studied recently, and showed that, 
in contrast to the Minkowski,  the de Sitter space is stable \cite{HWW}. 

With the above in mind, in  \cite{WWM} we studied perturbations of a scalar field cosmology.  
After deriving the generalized Klein-Gordon equation, { which is sixth-order} in
spatial derivatives,  we investigated scalar field perturbations coupled to gravity in a flat
Friedmann-Robertson-Walker (FRW) background. In the sub-horizon regime, we found that
in general the metric and scalar field modes have independent oscillations with different frequencies 
and phases. On super-horizon scales, the perturbations become adiabatic during
slow-roll inflation driven by a single field, and the comoving curvature perturbation is constant.

In this paper we shall generalize our previous studies \cite{WM} for scalar perturbations to
vector and tensor  perturbations in the SVW form of HL gravity \footnote{In \cite{KUY}, scalar perturbations in a flat FRW background 
were studied without fixing the gauge. This allowed the authors to study  properties of the gauge-invariant quantities 
easily.}. We will investigate
how standard results for linear perturbations in GR are modified and how it
may still be possible to recover some standard results in the long-wavelength or low-energy limit.
We will not consider the non-linear perturbations and consider only the linear evolution of perturbations
in a flat background. But, for vector perturbations, our results hold also for the non-flat cases.
%We implicitly assume that the strong-coupling (or ghost) problem can be
%addressed via this mechanism or some other approach.}
 
In Sec.~II we briefly review the HL cosmology in the SVW setup for
a flat background, while in Sec. III we present the general expressions for
vector perturbations, and show explicitly that they are the same
as those given in GR. We argue that this is also true even the 
background is not flat. In Sec. IV, we study the tensor perturbations
and present the general formulas, from which we find that high order 
derivatives of curvatures act as an anisotropic
stress, which could produce significant efforts on the high-frequency 
modes of gravitational waves. In Sec. V, as applications of our general
formulas for tensor perturbations, we study the power spectrum and
index in both the de Sitter and the power-law backgrounds, by using
the uniform approximations, proposed recently by {Habib} {\em et al}  \cite{Habib}.
When there is only one turning point, %$g(k, \eta)$ has only one zero, 
we obtain the analytical expressions
for the power spectrum and index in both cases. We conclude in Sec.~VI.

It should be noted that tensor perturbations were studied previously by several
authors in the framework of the HL theory.  
In particular, Takahashi and Soda studied the efforts
of primordial gravitational waves due to the parity violation \cite{TS1}, while Koh studied the
power spectrum and index of gravitational waves with the Corley-Jacobson 
dispersion relations \cite{Koh}. Yamamoto, Kobayashi and Nakamura, on the other
hand, studied the problems for both scalar and tensor perturbations, using the 
uniform approximations \cite{YKN}.    Gong, Koh and Sasaki  investigated  
vector and tensor perturbations in a different setup (In particular, the actions used by these authors violate the
parity) with a scalar field as the only source \cite{GKS}, and found 
that the vector perturbations have zero-degree of freedom, as that in GR.

%Myung studied the problem for the case  where the critical exponent $z =2$ \cite{Myung}. 

\section{The Flat FRW Background in the SVW Setup}
\renewcommand{\theequation}{2.\arabic{equation}} \setcounter{equation}{0}

The SVW generalization \cite{SVW} of HL theory coupled with matter fields has been reviewed in our previous
work \cite{WM,WWM}, and in this paper we shall directly adopt the notations and conventions given 
there without further explanations. For detail, we refer readers to \cite{WM,WWM}.

The flat homogeneous and isotropic universe is described by the metric, $ds^{2} = a^{2}(\eta)\left(- d\eta^{2} 
+ \delta_{ij}dx^{i}dx^{j}\right)$. For this metric, $\bar K_{ij} = - a {\cal{H}} \delta_{ij}$,  where
$ {\cal{H}} = {a'}/a$ and a prime denotes derivative with respect to $\eta$.   Then, it can be shown that
the Hamiltonian constraint, 
 \bq \lb{eq1}
\int{ d^{3}x\sqrt{g}\left({\cal{L}}_{K} + {\cal{L}}_{{V}}\right)}
= 8\pi G \int d^{3}x {\sqrt{g}\, J^{t}},
 \eq
 yields the (generalized) Friedmann equation,
 \bq
 \lb{2.1}
 \Big(1 - \frac{3}{2}\xi\Big)\frac{{\cal{H}}^{2}}{a^{2}} = \frac{8\pi G}{3}\bar{\rho} + \frac{\Lambda}{3},
 \eq
 while the dynamical equations,
  \bqn \lb{eq3}
&&
\frac{1}{N\sqrt{g}}\left(\sqrt{g}\pi^{ij}\right)'%^{\displaystyle{\cdot}}
= -2\left(K^{2}\right)^{ij}+2 \left(1 - \xi\right)K K^{ij}
\nb\\
& &~~ + \frac{1}{N}\nabla_{k}\left[N^k \pi^{ij}-2\pi^{k(i}N^{j)}\right] \nb\\
& &~~ + \frac{1}{2} {\cal{L}}_{K}g^{ij}   + F^{ij} + 8\pi G
\tau^{ij},
 \eqn
give rise to
\bq
\lb{2.2}
 \Big(1 - \frac{3}{2}\xi\Big)\frac{2{\cal{H}}' + {\cal{H}}^{2}}{a^{2}}    =  - {8\pi G}\bar{p} + {\Lambda},
\eq 
where a prime denotes derivative with respect to $\eta$, 
 \bq\label{mq}
\bar{J}^t=-2\bar\rho,~~ \bar{J}^i=0,~~ \bar \tau_{ij} = a^{2}\bar p\,
\delta_{ij},
 \eq
and $\bar\rho$ and $\bar p$ are the total density and pressure. Similar to that in GR, the momentum 
constraint,
 \bq \lb{eq2}
\nabla_{j}\pi^{ij} = 8\pi G J^{i},
 \eq
 is satisfied identically, while the conservation law of energy 
  \bq \lb{eq4a} 
 \int d^{3}x \sqrt{g} { \left[{g}_{kl}'\tau^{kl} -
 \frac{1}{\sqrt{g}}\left(\sqrt{g}J^{t}\right)'%^{\displaystyle{\cdot}}
 % \right.  }  % \nb\\ & &  \left.  \;\;\;\;\;\;\;\;\;\;\;\;\;\;\;\;\;\; 
+  \frac{2N_{k}}
 {N\sqrt{g}}\left(\sqrt{g}J^{k}\right)'%^{\displaystyle{\cdot}}
 \right]} = 0,
 \eq
 yields 
  \bq \lb{2.3}
{\bar{\rho}}' + 3{\cal{H}} \left(\bar\rho +\bar p \right) = 0.
 \eq
For the FRW background, the conservation law of momentum 
\bqn
\lb{eq4b}  
\nabla^{k}\tau_{ik} &-&
\frac{1}{N\sqrt{g}}\left(\sqrt{g}J_{i}\right)' %^{\displaystyle{\cdot}}
 - \frac{N_{i}}{N}\nabla_{k}J^{k} \nb\\
&- &   \frac{J^{k}}{N}\left(\nabla_{k}N_{i}
- \nabla_{i}N_{k}\right) =
 0,
\eqn
is satisfied identically. 

It should be noted that Eq. (\ref{2.3}) can be also obtained directly from Eqs. (\ref{2.1}) and (\ref{2.2}).
In addition, replacing $G$ and $\Lambda$ by $G/(1 - 3\xi/2)$ and $\Lambda/(1 - 3\xi/2)$,  Eqs. (\ref{2.1}) 
and (\ref{2.2}) becomes identical to those given in GR.

%\subsection{Vector perturbations}
\section{Cosmological vector perturbations}

\renewcommand{\theequation}{3.\arabic{equation}} \setcounter{equation}{0}

The cosmological vector perturbations of the metric are given by   \cite{WM}
 \bqn \lb{2.4}
\delta{g}_{ij} &=&   a^{2}(\eta)\big( F_{i,j} + F_{j,i}\big), %h_{ij}\left(\eta, x^{k}\right)
 \nb\\
\delta{N}^{i} &=&    - S^{i},~~~\delta{N} = 0,
%a(\eta) n(\eta) ,
 \eqn
where
% \bq
%\lb{2.5}
%h_{ij} =   F_{i,j} + F_{j,i},\;\;\;  n_{i} =  - S_{i}, 
% \eq
%with 
$F_{i,j} \equiv \partial F_{i}/\partial x^{j}$ and 
 \bq \lb{2.6}
S_{i}^{\;\;, i} = 0,\;\;\; F_{i}^{\;\;,i} = 0,
\eq
with $S_{j}^{\;\;, i} \equiv \delta^{ik}S_{j,k}$.  
The corresponding matter perturbations are given by
 \bq
 \lb{2.7}
\delta{J}^{i} = \frac{1}{a^{2}}q^i, \; \delta J^{t} = 0,\;
 \delta{\tau}^{ij} = \frac{2}{a^{2}} \Big(\Pi^{(i,j)} - \bar{p}{{F}}^{(i,j)}\Big),
 \eq
where $ %{\cal{F}}^{ij} \equiv \big(F^{i,j} + F^{j,i}\big)/2,\;
 f^{(ij)} \equiv \big(f^{ij} + f^{ji}\big)/2$ and 
\bq
\lb{2.7a}
q^i{}_{,i}=0 = \Pi^{i}_{\;\;,i}. 
\eq
Note the slight difference between the definition of $ \delta{\tau}^{ij}$ used here and the one introduced in
\cite{WM}. The vectors 
$S^{i}, \; F^{i}, \; q^{i}$ and $\Pi^{i}$ are in general functions of $\eta$ and $x^i$, and all their 
indices are  lowered by $\delta_{ij}$, for example,  $S_{i} \equiv \delta_{ik}S^{k}$  and so on. 
With the quasi-longitudinal gauge, one can set $F_{i} = 0$. However, to have our results as much applicable
as possible, we shall leave this possibility open, and consider the case with any $F_i$ and $S_i$.
%, a gauge that will be used in this paper.
Then, we find that
\bqn
\lb{2.8}
K_{ij} &=& - a\Big({\cal{H}} \delta_{ij} + {{F}}_{(i,j)}' + 2{\cal{H}}  {{F}}_{(i,j)} +  {{S}}_{(i,j)}\Big),\;\;\; R_{ij} = 0, \nb\\
{\cal{L}}_{K} &=& = - \frac{3(2-3\xi)}{a^{2}} {\cal{H}}^{2}, \;\;\; {\cal{L}}_{V} = 2\Lambda,\nb\\
F^{ij} &=& \sum^{8}_{s=0}{g_{s}\zeta^{n_{s}}\left(F_{s}\right)^{ij} }= - \frac{\Lambda}{a^{2}}\Big(\delta^{ij}
- 2{{F}}^{(i,j)}\Big).
\eqn
%where
%\bq
%\lb{2.9}
%{\cal{S}}_{ij} \equiv \frac{1}{2}\big(S_{i,j} + S_{j,i}\big).
%\eq
Hence,  the Hamiltonian constraint (\ref{eq1})   yields the generalized Friedmann equation (\ref{2.1}),
 while the momentum constraint (\ref{eq2}) gives
 \bq
 \lb{2.10}
 \partial^{2}\Big({{F}}_{i}' + {{S}}_{i}\Big) = 16\pi G a q_{i},
 \eq
 where $ \partial^{2} \equiv \delta^{ij}\partial_{i}\partial_{j}$. It is interesting to note that the quantity,
\bq
\lb{2.11a}
\Phi_{i} \equiv {{F}}_{i}' + {{S}}_{i},
\eq
is gauge-invariant \cite{WM}. 
The dynamical equation (\ref{eq3}), on the other hand, yields Eq. (\ref{2.2}) to zeroth order, while to first order, it 
gives
\bq
\lb{2.11}
\Big({{F}}_{(i,j)}' + {{S}}_{(i,j)}\Big)'  + 2{\cal{H}}\Big({{F}}_{(i,j)}' + {{S}}_{(i,j)}\Big) = 16\pi G a^{2} \Pi_{(i,j)}.
\eq
The conservation law of energy (\ref{eq4a}) does not give new constraint, rather than Eq. (\ref{2.3}), while the 
conservation of momentum yields,
\bq
\lb{2.12}
q_{i}' + 3{\cal{H}} q_{i} = a \partial^{2}\Pi_{i}.
\eq
However, this equation is not independent, and can be obtained from Eqs. (\ref{2.10}) and (\ref{2.11}). 

It is remarkable that neither the parameter $\xi$ nor high order derivatives are involved in Eqs. (\ref{2.10}) - (\ref{2.12}).
The reasons are the following: All places that depend on $\xi$ are through the term $(1-\xi)K$, as one can see from
the definitions of ${\cal{L}}_{K}, \; \pi^{ij}$ and the dynamical equations Eq. (\ref{eq3}).  However, to first order, from
Eq. (\ref{2.8}) we find that $K = -3{\cal{H}}/a$. Thus, the linear perturbations of $K$ vanish identically.  Then, all
equations of linear perturbations do not depend explicitly on $\xi$. On the other hand, all coupling constants
$g_{2}, \; g_{3}, ..., g_{8}$ are proportional to high order derivatives through the 3-dimensional Ricci tensor $R_{ij}$ 
and its derivatives. Since $R_{ij}$ also vanishes identically even when $F_{i} \not= 0$,  these high order derivatives have
no contributions to the linearized equations.  %and in particular these coupling constants do not appear
%in the equations for $S_{i}$. 
%This  explains why high order derivatives of $F_{i}$ and $S_{i}$ do not appear in the equations. 
%Note that these arguments are also true for the non-flat FRW models. To see this clearly, one may first choose
%the quasi-longitodinal gauge in which $F_{i} = 0$. Clearly $R_{ij}$ and its derivatives will hot depend on $S_{i}$.
%Then, one can come back to the original coordinates. Since 
As a result, Eqs. (\ref{2.10})-(\ref{2.12})  do not depend on $\xi, \; g_{2}, ..., g_{8}$, and
are identical to those given in GR \cite{MW09}, by noticing
\bq
\lb{2.13}
q_{i} = - a\delta{q}_{i} = a\big(\bar{\rho} + \bar{p}\big)\big(S_{i} - v_{i}\big),
\eq
where $\delta{q}_{i}$ and $v_{i}$ are quantities introduced in \cite{MW09}. 

The above conclusion can be further generalized 
to the non-flat case. To see this, the simplest way is
to work with the gauge  $F_{i} = 0$ \cite{WM}, so
the gauge-invariant vector defined by Eq. (\ref{2.11a}) reduces exactly 
to $S_{i}$. In this gauge, since $\delta R_{ij} = 0$, one can see that the high order derivatives of
curvature have no contributions. In addition,  to first  order we also have  $\delta K = 0$. Then, as argued above, the parameter
$\xi$ will not appear in the linearized equations. Therefore,
% Then, using the same arguments given above  one can easily show that
 {\em the resulting linearized equations for $S_{i}$ do not depend explicitly on $\xi, \; g_{2}, ..., g_{8}$ even  
 when the spatial curvature of the 
FRW universe is different from zero, and must be identical to those given in
GR} \cite{MW09}.  Hence, the results obtained in GR
can be easily generalized to the HL theory. For example, for a scalar field, both $q_{i}$ and $\Pi_{i}$ vanish identically
in the HL theory \cite{WWM}. Then, similar to that in GR, the vector perturbations are zero in all of space and shall remain
so, if no sources of vorticity are introduced. 

%{\bf Note that $\Pi_{i}$ used in this paper is half of that introduced by David and Karim in \cite{MW09}, i.e.,
%$\Pi_{i} = \Pi^{MW}_{i}/2$. Then, it seems that
%a factor $1/2$ is missing in the right-hand side of Eq.(8.63) in \cite{MW09}. On the other hand, the factor $a^{3}2\pi G$
%in the right-hand side of Eq.(2.92) in Karim's paper, arXiv:astro-ph/0101563, should be replaced by $a^{3}4\pi G$.
%David, could you please confirm this? }

\section{Cosmological tensor perturbations}

\renewcommand{\theequation}{4.\arabic{equation}} \setcounter{equation}{0}

The cosmological tensor perturbations of the metric are given by   \cite{WM}
 \bq \lb{3.1}
\delta{g}_{ij} =   a^{2}(\eta)H_{ij}\left(\eta, x^{k}\right),\;\;\;
\delta{N}^{i} =     0 =  \delta{N},
%a(\eta) n(\eta) ,
 \eq
with  the constraints
 \bq 
\lb{3.2}
 H^{i}_{i}  = 0=  H_{ij}^{\;\;\;,j},
 \eq
 while the corresponding matter perturbations are given by 
  \bq
 \lb{3.3}
 \delta{\tau}^{ij} = \frac{1}{a^{2}}  \left(\Pi^{(TT) ij} - \bar{p}H^{ij}\right) , \;\;\; \delta{J}^{i} =  0 = \delta J^{t},
  \eq
where
\bq
\lb{3.4}
{\Pi^{(TT) i}}_{ i} = 0 = {\Pi^{(TT) ij}}_{,j},
\eq
and $ \Pi^{(TT) ij} = \Pi^{(TT) ij}\left(\eta, x^{k}\right)$. Note the difference between $\Pi^{(TT) ij}$ used here
and $\Pi^{ ij}$ defined in \cite{WM}. All  indices of $H_{ij}$ and $\Pi^{(TT)}_{ij}$ will be raised by
$\delta^{ij}$. Then, we find that  to first-order the extrinsic curvature and Ricci tensors are given by
\bqn
\lb{3.5}
K_{ij} &=& - a {\cal{H}}\delta_{ij} - \frac{a}{2}\big(H_{ij}' + 2{\cal{H}}H_{ij}\big),\nb\\
R_{ij} &=& - \frac{1}{2} \partial^{2}H_{ij}.
\eqn
 Because of the constraints (\ref{3.2}), it can be shown that
in the present case the first-order perturbations of ${\cal{L}}_{K}$ and ${\cal{L}}_{V}$ vanishes identically, 
\bqn
\lb{3.6}
{\cal{L}}_{K} &=& - \frac{3(2-3\xi)}{a^{2}}{\cal{H}}^{2} + {\cal{O}}\big(H^{2}\big),\nb\\
{\cal{L}}_{V} &=& 2 \Lambda + {\cal{O}}\big(H^{2}\big).
\eqn
As a result, to zeroth-order the Hamiltonian constraint (\ref{eq1}) yields the Friedmann equation, while to first-order it is
satisfied identically. On the other hand, from the expression,
\bq
\lb{3.7}
\pi^{ij} = - \frac{2-3\xi}{a^{3}}{\cal{H}}\delta^{ij} +  \frac{1}{2a^{3}}\Big[{H^{ij}}' + 2(2-3\xi){\cal{H}}H^{ij}\Big],
\eq
we find that the momentum constraint (\ref{eq2}) is also satisfied identically for tensor perturbations, where $\delta{J}^{i} =  0$. 
This is also true for the   momentum conservation   (\ref{eq4b}), while    the energy   conservation (\ref{eq4a}) yields Eq. (\ref{2.3})
to zero-th order, and is identically satisfied to first-order. %On the other hand,  
To first-order we also find that
\bqn
\lb{3.8}
F^{ij} &\equiv& - \frac{1}{\sqrt{g}}\frac{\delta\big(\sqrt{g}{\cal{L}}_{V}\big)}{\delta{g}_{ij}} 
     = \sum_{s=0}^{8}{g_{s}\zeta^{n_{s}}\left(F_{s}\right)^{ij}} =   - \frac{\Lambda}{a^{2}} \delta^{ij} \nb\\
     &&  + \frac{1}{2a^{2}}\left(2\Lambda + \frac{1}{a^{2}}\partial^{2} 
           - \frac{g_{3}}{\zeta^{2}a^{4}}\partial^{4}  - \frac{g_{8}}{\zeta^{4}a^{6}}\partial^{6}\right)H^{ij}. \nb\\
\eqn  
Then, the dynamical equations (\ref{eq3}) yield,
%\bqn
%\lb{3.8}
%H_{ij}'' &+& 2 {\cal{H}}H_{ij}' + (2-3\xi)\big(2{\cal{H}}' + {\cal{H}}^{2}\big)H_{ij} = 16\pi G a^{2}\Pi_{ij}\nb\\
%&+&  a^{2}\Big(2\Lambda + \frac{1}{a^{2}}\partial^{2} 
%           - \frac{g_{3}}{\zeta^{2}a^{4}}\partial^{4}  - \frac{g_{8}}{\zeta^{4}a^{6}}\partial^{6}\Big)H_{ij}.\nb\\
%\eqn
%Setting
%\bq
%\lb{3.9}
%\Pi_{ij} = \frac{1}{2}\Pi^{(TT)}_{ij} - \bar{p}H_{ij},
%\eq
%where $\Pi^{(TT)}_{ij}$ satisfies  the same  constraints (\ref{3.4}), the above equation can be written as
\bqn
\lb{3.10}
H_{ij}'' &+& 2 {\cal{H}}H_{ij}' - \partial^{2}H_{ij} =  16\pi G a^{2}\Pi^{(TT)}_{ij}\nb\\
&- &  \frac{1}{\zeta^{2}a^{2}}\Big({g_{3}}+ \frac{g_{8}}{\zeta^{2}a^{2}}\partial^{2}\Big)\partial^{4}H_{ij}.
\eqn
Note that in writing the above equation, we had used Eq. (\ref{2.2}). In addition, 
 the Newtonian constant $G$ is not modified by the factor $1/(1-3\xi/2)$, as it was
 for scalar perturbations \cite{WM}. 
When $g_{3} = g_{8} = 0$, it reduces exactly to that given in GR \cite{MW09}. 
When they are different from zero, it shows clearly that these high order derivatives serve as
anisotropic sources to produce primordial gravitational waves.  However, since 
$\zeta^{2} = M^{2}_{pl}/2$, they are highly suppressed in the IR regime. 

During inflation we can neglect $\Pi^{(TT)}_{ij} = 0$, 
since the inflaton has no anisotropic stress. Then, the effective gravitational  stress,
\bq
\lb{3.14}
\Pi^{HL}_{ij} \equiv - \frac{1}{16\pi G \zeta^{2}a^{4}}\Big(g_{3} +  \frac{g_{8}}{\zeta^{2}a^{2}}\partial^{2}\Big)\partial^{4}H_{ij},
\eq
affects only the high-frequency modes, which could be very interesting, as they provide a mechanism to produce 
the initial seeds of gravitational waves even during the epoch of inflation.

Introducing two eigenmodes $e^{(+, \; \times)}_{ij}(x)$   of the spatial Laplacian,
$\big(\partial^{2} + k^{2}/a^{2}\big)  e^{(+, \; \times)}_{ij}(x) = 0$ with comoving wavenumber $k$, we can
decompose $H_{ij}$ and $\Pi^{(TT)}_{ij}$  into two independent components:
\bqn
\lb{3.11}
H_{ij}(\eta, x) &=& H_{(+, \times)}(\eta)e^{(+, \; \times)}_{ij}(x),\nb\\
\Pi^{(TT)}_{ij}(\eta, x)  &=& \Pi^{(TT)}_{(+,\times)}(\eta)e^{(+, \; \times)}_{ij}(x),
\eqn
where $e^{(+, \; \times)}_{ij}$ denote two possible polarization states of gravitational waves, $+$ and $\times$. 
Then, Eq. (\ref{3.10})
reduces to
\bq
\lb{3.12}
w_{k}'' + \Big(\omega^{2}_{T} - \frac{a''}{a}\Big) w_{k} =  16\pi G a^{3} \Pi^{(TT)},
\eq
where $w_{k} = \big(w^{+}_{k}, w^{\times}_{k}\big)$ etc.,  and
\bqn
\lb{3.13}
w^{(+,\; \times)}_{k} &=& a H^{(+, \times)}_{k},\nb\\
 \omega^{2}_{T} &\equiv& k^{2} +  \frac{{g_{3}}k^{4}}{\zeta^{2}a^{2}} - \frac{g_{8}k^{6}}{\zeta^{4}a^{4}}.
 \eqn
Note that in writing Eq. (\ref{3.12}) we dropped the sub-indices $+$ and $\times$ from  $w$ and $\Pi^{(TT)}$. 
In the UV regime, $ \omega^{2}_{T}  \simeq -g_{8}k^{6}/(\zeta^{4}a^{4})$, and to have  stable modes we must
assume  
\bq
\lb{3.13a}
g_{8} <0. 
\eq
Then, the primordial gravitational wave spectra are  scale-invariant \cite{Muk,TS1}. In the 
IR regime, $ \omega^{2}_{T}  \simeq  k^{2}$, %whereby the GR limit is recovered, 
and   the $H = Const$ 
mode on large scales is regained.  Since the intermediate $k^4$ 
part is not scale-invariant, there may be a peak in the spectra. In addition,  both states  $+$ and $\times$ satisfy the same 
equation, in contrast to the case with detailed balance condition \cite{TS1}, circular polarization cannot be 
generalized when $\Pi^{(TT)} = 0$ in the current setup. This is because the SVW generalization preserves parity \cite{SVW}, while 
the HL theory with detailed balance condition does not \cite{TS1}.    

%In addition, % After we started to write up this paper, we also became aware of the work of 
%Gong, Koh and Sasaki also investigated  
%vector and tensor perturbations in a different setup (In particular, the actions used by them violate the
%parity) with a scalar field as the only source \cite{GKS}. It was also found that the vector perturbations 
%have zero-degree of freedom, as that in GR.
 
It is also interesting to note that, in contrast to scalar perturbations \cite{WM}, Eq. (\ref{3.10}) does not
contain the coupling constant $\xi$ explicitly, with the same reason as for vector perturbations as explained 
following Eq. (\ref{2.12}). This is also true  for  non-flat FRW models, because the kinetic part is independent
of the spatial curvature. Then, {\em tensor perturbations in all FRW models does not have the ghost problem for
any given coupling constant $\xi$},  including the range $0 \le \xi \le 2/3$, in which ghosts were found in  
the scalar sector of perturbations \cite{WM,KA}.

\section{Tensor perturbations in some specific backgrounds}

\renewcommand{\theequation}{5.\arabic{equation}} \setcounter{equation}{0}

 In this section, we consider tensor perturbations of Eq. (\ref{3.12}) with the assumption
 that the efforts of $\Pi^{(TT)}$ are negligible, so 
 Eq. (\ref{3.12}) reduces to
 \bq
\lb{3.14a}
w_{k}'' + \Big(\omega^{2}_{T} - \frac{a''}{a}\Big) w_{k} = 0.
\eq
The above equations usually has the following asymptotic solutions,
\bq
\lb{3.15}
w_{k} = \cases{\frac{w^{0}_{k}}{\sqrt{2\omega_{T}}}e^{- i\omega_{T}\eta}, & $ k\eta \rightarrow - \infty$,\cr
A_{k}a, & $ k\eta \rightarrow 0^{-}$,}
\eq
where $w^{0}_{k}$ and $A_{k}$ are constants. Then, in the superhorizon region ($k\eta \simeq 0$), the power spectrum is given by
\bq
\lb{3.16}
\left. P_{T}(k)\right|_{k\eta  \rightarrow 0^{-}} = \left. \frac{k^{3}}{2\pi^{2}}\left|\frac{w_{k}}{a}\right|^{2}  \right|_{k\eta \rightarrow 0^{-}}
=   \frac{k^{3}}{2\pi^{2}}\left|A_{k}\right|^{2}.
\eq
Therefore, to find $ P_{T}(k)$ now simply reduces to find the constant $A_{k}$ by connecting the two asymptotic solutions. This
can be done by a matching process in the intermediate region [cf. Fig. \ref{fig1}] 
\cite{SL93}. As pointed out by several authors
\cite{WMS,Stewart}, such a process lacks error control and is not systematically improvable.  

%%%%%%%%%%%%%%%%%%%%%%%%%%%%%%%%%%%%%%%%%%%%%%%%%%%%%%%%%%%%%%%%%%%%%%%%%%%%%%%
\begin{figure}
\includegraphics[width=\columnwidth]{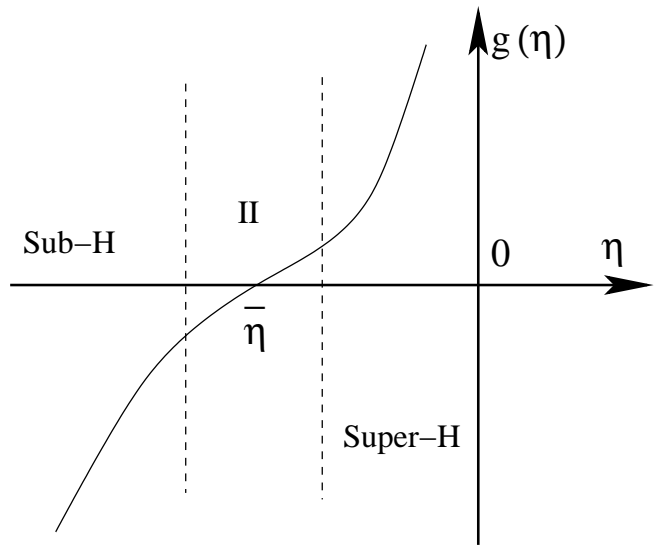}
\caption{The three different regions: (a) the sub-horizon region ($k\eta \rightarrow - \infty$), denoted by Sub-H;
(b) the imtermediate region ($\eta \simeq \bar{\eta}$), denoted by $II$; and (c) the super-horizon region ($k\eta \rightarrow 0^{-}$), denoted by Super-H. 
For well-defined $\omega_{T}$, the function $g(\eta)$ is positive for $\eta > \bar{\eta}$, and  negative  for $\eta < \bar{\eta}$, 
where $\bar{\eta}$ is called the turning point, and given by the negative root of $g(\bar{\eta}) = 0$.}
\label{fig1}
\end{figure} 
%%%%%%%%%%%%%%%%%%%%%%%%%%%%%%%%%%%%%%%%%%%%%%%%%%%%%%%%%%%%%%%%%%%%%%%%%%%%%%%

Recently, Habib {\em et al} \cite{Habib} advocated another method - the so-called uniform approximation \cite{Olver}. The latter provides 
a single approximate solution for the whole range $k$, so it does not employ the intermediate matching process and the approximation 
procedure can be systematically improved and possesses an error control function. To show how this method works, we first write 
Eq. (\ref{3.14a}) in the form \cite{Habib,YKN},
 \bq
 \lb{3.17}
 w_{k}''  = \Big[g(k, \eta) + q(\eta)\Big]w_{k},
 \eq
 where
 \bqn
 \lb{3.18}
 g(k,\eta) &\equiv& \frac{a''(\eta)}{a(\eta)} + \frac{1}{4\eta^{2}} - \omega_{T}^{2}(k,\eta), \nb\\
 q(\eta) &\equiv& - \frac{1}{4\eta^{2}}.
 \eqn
%Note that the function $g(\eta)$ also depends  on $k$, i.e., $g(\eta) = g(k, \eta)$. In addition, 
Note that the specific choice of the  function $q(\eta)$  is to guarantee the convergence of the approximation \cite{Olver}. Then, the single 
 approximate solution can be written as
 \bq
 \lb{3.19}
 w_{k} = \left(\frac{y(k, \eta)}{g(k, \eta)}\right)^{1/4} \Big[a_{k} Ai(y) + b_{k} Bi(y)\Big],
 \eq	
 where $Ai(y)$ and $ Bi(y)$ are Airy functions, and 
 \bq
 \lb{3.20}
 y(k, \eta) = \cases{y_{+}(k, \eta), &$ \eta > \bar{\eta}$,\cr
 y_{-}(k, \eta), & $ \eta < \bar{\eta}$,\cr}
 \eq
 with 
 \bq
 \lb{3.21}
 y_{\pm}(k, \eta) = \pm \Bigg\{\pm \frac{3}{2} \int_{\bar{\eta}(k)}^{\eta}{\sqrt{\pm g(k, \eta')} d\eta'}\Bigg\}^{2/3},
 \eq
 and $\bar{\eta}$ is the turning point, defined by $g(k, \bar{\eta}) = 0$ [cf. Fig. \ref{fig1}]. %, which in general is function of $k$,
% i.e., $\bar{\eta} = \bar{\eta}(k)$. 
The integration of $y_{-}$ is taken on the left of the turning point $\bar{\eta}$, while 
the one of $y_{+}$ is taken on the right of  $\bar{\eta}$.
%  Then, the error
 %control function $\epsilon(k, \eta)$ is given by 
% \bqn
% \lb{3.22}
% \epsilon(k, \eta) &=& \pm \frac{1}{4} \int^{\bar{\eta}(k)}_{\eta}{\big(\pm g\big)^{-3/2}\left(g'' - \frac{5{g'}^{2}}{4g} - \frac{g}{\eta^{2}}\right)d\eta}\nb\\
% & & 
% \pm \frac{5}{24\left|y_{\pm}\right|^{3/2}},
% \eqn
% where the upper (lower) sign is taken for $\eta > \bar{\eta}(k) \; (\eta < \bar{\eta}(k))$.
 In order to fix the coefficients $a_{k}$ and $b_{k}$, $w_{k}$ is required to reduce to its asymptotic form (\ref{3.15}) 
 as $k \eta \rightarrow - \infty$. In this limit, for well-behavior $\omega_{T}$, the function $y_{-}(k, \eta)$ is very large and
 negative. So, one can use the asymptotic forms of the Airy functions \cite{AS72},
 \bqn
 \lb{3.23}
 Ai(-x) & \simeq & \frac{1}{\left(\pi^{2} x\right)^{1/4}} \sin\left(\frac{2}{3}x^{3/2} + \frac{\pi}{4}\right),\nb\\
 Bi(-x) &\simeq &  \frac{1}{\left(\pi^{2} x\right)^{1/4}} \cos\left(\frac{2}{3}x^{3/2} + \frac{\pi}{4}\right),
 \eqn
for $x \gg 1$ with $\left|{\mbox{arg}} (x) \right| < 2\pi/3$. Then, choosing
\bq
\lb{3.25}
a_{k} = -i \sqrt{\frac{\pi}{2}} \; e^{i\pi/4}, \;\;\;  b_{k} =  \sqrt{\frac{\pi}{2}} \; e^{i\pi/4},
\eq
we find that 
\bq
\lb{3.26}
w_{k} = \frac{1}{\sqrt{2\omega_{T}}} \exp\Bigg\{ - i  \int_{\eta}^{\bar{\eta}(k)}{\omega_{T}(k, \eta')d\eta'}\Bigg\},
\eq 
as $\eta \rightarrow - \infty$, which  is exactly the required adiabatic form  in the sub-horizon region
($ - k \eta  \gg 1$).  It can be shown that the Wronskian normalization condition,
\bq
\lb{3.24}
w_{k}{w_{k}^{*}}' - {w_{k}}' w_{k}^{*} = i,
\eq
is satisfied for the choice of Eq. (\ref{3.25}).

To find the asymptotic behavior of the solution (\ref{3.19}) for $k \eta \rightarrow 0^{-}$, we first notice that \cite{AS72},
 \bqn
 \lb{3.27}
 Ai(x) &\simeq& \frac{1}{2\left(\pi^{2} x\right)^{1/4}} e^{-\frac{2}{3}x^{3/2}},\;\;  \big(\left|{\mbox{arg}}(x)\right| < \pi\big), \nb\\
 Bi(x) &\simeq&  \frac{1}{\left(\pi^{2} x\right)^{1/4}} e^{\frac{2}{3}x^{3/2}},\;\;   \big(\left|{\mbox{arg}}(x)\right| < \pi/3\big), ~~
 \eqn
as $x  \rightarrow \infty$. % with $\left| {\mbox{arg}}(x)\right| < \pi/3$. 
  Then, Eq. (\ref{3.19}) has the asymptotics,
 \bq
 \lb{3.28}
 w_{k} \simeq  \frac{e^{i\pi/4}}{\left[4 g(k,\eta)\right]^{1/4}} %\exp\Bigg\{\int^{\eta}_{\bar{\eta}(k)}{\sqrt{g(k, \eta')}d\eta'}\Bigg\},
 \exp\big[{\cal{D}}(k, \eta)\big],
 \eq
 as $ k\eta \rightarrow 0^{-}$, where
 \bq
 \lb{3.30a}
 {\cal{D}}({k,\eta}) \equiv \int^{\eta}_{\bar{\eta}(k)}{{\sqrt{g(k, \eta')}{d\eta'}}}.
 \eq
Hence,   we find that
 \bq
 \lb{3.29}
 P_{T}(k) = \lim_{k\eta \rightarrow 0^{-}}\frac{k^{3}}{4\pi^{2}a^{2}\sqrt{g(k, \eta)}}
  \exp\Big\{2{\cal{D}}(k, \eta)\Big\},
  \eq
 and  the spectrum index of the quantum fluctuations is given by
 \bq
 \lb{3.30}
 n_{T} \equiv  \left. \frac{d\ln{P_{T}}}{d\ln{k}} \right|_{k\eta \rightarrow 0^{-}}
          = 3 + \lim_{k\eta \rightarrow 0^{-}}  2k\frac{d{\cal{D}}(k, \eta)}{dk}.
\eq
Note that in writing down the above expression, we had assumed that
\bq
\lb{3.0b}
 \lim_{k\eta \rightarrow 0^{-}}  \frac{kg_{,k}(k, \eta)}{g(k,\eta)} = 0.
 \eq
 It is interesting to note that both $ P_{T}(k) $ and $n_{T}$ are uniquely determined by the function
 ${\cal{D}}({k,\eta})$.  % as can be seen clearly from Eqs. (\ref{3.21}),  (\ref{3.29}) and  (\ref{3.30}). 
 %, where the integration   is taken on the right of the turning point  $\bar{\eta}$. % of the turning point $\bar{\eta}$.
 It is also important to note that the above formulas are valid for any $\omega_{T}(k,\eta)$, as long as
  $g(k,\bar{\eta}) = 0$ has only one negative root. When $g(k, \bar{\eta})$ has multiple zeros for $\eta < 0$, different
 treatment \cite{Olver} is needed. In this paper, we shall consider only the case where   $g(k,\bar{\eta}) = 0$
 has only one negative root, as shown in Fig. \ref{fig1}. In particular,  %In the rest of this section, 
 we shall consider two explicit backgrounds: the de Sitter spacetime, and the 
 spacetime with power-law expansion.

 \subsection{The de Sitter Background}
 
 In the de Sitter background, we have $a(\eta) = -1/(H\eta)$, and  
 \bq
\lb{3.31} 
 \omega^{2}_{T} =  k^{2} +  \frac{g_{3}H^{2}k^{4}}{\zeta^{2}} \eta^{2} 
 - \frac{g_{8}H^{4}k^{6}}{\zeta^{4}} \eta^{4}.
 \eq
Before studying the above equation, it is interesting to compare it with the one obtained in \cite{TS1}
with detailed balance condition, for which the tensor perturbations can be also cast in the form of
Eq. (\ref{3.13}) but now with a different $\omega^{2}_{T}$, given by
\bq
\lb{3.32}
\omega^{2}_{T,TS} =  c^{2}k^{2}\Bigg[1  + \beta\big(ck\eta\big)^{2}\Big(1 + c\epsilon^{A} \gamma k \eta\Big)^{2}\Bigg],
\eq
where $c$ is the ``emergent speed" of light, and
\bq%n
\lb{3.33}
c^{2} %&\equiv&
\equiv  \frac{\kappa^{4}\mu^{2}\Lambda_{w}}{16(1- 3\lambda)},\;\;%\nb\\
\beta %&\equiv& 
\equiv \frac{(1 - 3\lambda) H^{2}}{c^{2}\Lambda_{w}}, \;\;
\gamma \equiv \frac{2H}{c\mu w^{2}},
\eq%n
and $\lambda \equiv 1 - \xi$. The constants $\mu,\; w$ and $\Lambda_{w}$ are the free parameters in the model,
and $\epsilon^{A} = \pm 1$. When $\epsilon^{A} = 1$, it is called the right-handed mode, and  when $\epsilon^{A} =  - 1$ 
it is called  the left-handed mode.  As mentioned above, the difference between left- and right- handed is exactly due to
 the violation of the parity. 

In the SVW setup, the theory is explicitly parity-preserving, so the right- and left-handed modes satisfy the same
equation, and $\omega^{2}_{T}$ does not depend on $\epsilon^{A}$, and is given by Eq. (\ref{3.31}). Then, from
Eq. (\ref{3.18}) we find that
\bq
\lb{3.34}
g(k, \eta) %= \frac{ 9}{4\eta^{2}} - \omega_{T}^{2} %\nb\\ &=&  
= k^{2}\Bigg[\frac{ 9}{4z^{2}} - \Big(1 + \tilde{g}_{3} z^{2}
                    +  \tilde{g}_{8}z^{4}\Big)\Bigg],
 \eq
where
\bq
\lb{3.35}
z \equiv k\eta,\;\;\; \tilde{g}_{3} \equiv \frac{g_{3}H^{2}}{\zeta^{2}}, \;\;\;
\tilde{g}_{8} \equiv \frac{|g_{8}|H^{4}}{\zeta^{4}} > 0.
\eq
It can be shown that the function $g(k, \eta)$ defined above satisfies the condition (\ref{3.0b}).

Let first consider the case $g_{3} = g_{8} = 0$, for which the turning point is at 
\bq
\lb{3.36}
\bar{\eta} = - \frac{3}{2k},
\eq
and Eq. (\ref{3.30a}) yields
\bq
\lb{3.37}
{\cal{D}}(k, \eta) = - \left\{\sqrt{\frac{9}{4} - k^{2}\eta^{2}} + \frac{3}{2} \ln\frac{\big(-k\eta\big)}{\frac{3}{2} + \sqrt{\frac{9}{4} - k^{2}\eta^{2}}}\right\}.
\eq
 Then, from Eqs. (\ref{3.29}) and (\ref{3.30}) we obtain
\bq
\lb{3.38}
P_{T}(k) = \frac{9H^{2}}{2\pi^{2} e^{3}},\;\;\; n_{T} = 0,
\eq
which are the well-known results obtained in GR \cite{Lid97}.

 %%%%%%%%%%%%%%%%%%%%%%%%%%%%%%%%%%%%%%%%%%%%%%%%%%%%%%%%%%%%%%%%%%%%%%%%%%%%%%%
\begin{figure}
\includegraphics[width=\columnwidth]{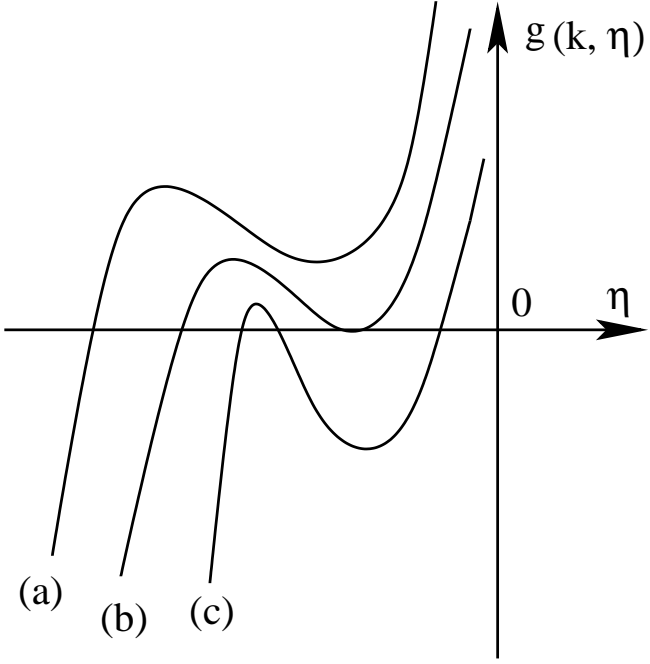}
\caption{The three different cases for the function $g(k, \eta)$ defined by Eq. (\ref{3.34}): (a)  $g(k, \eta)$ has one turing
point;  (b) $g(k, \eta)$ has two turing points; and   (c) $g(k, \eta)$ has three turing points.}
\label{fig2}
\end{figure} 
%%%%%%%%%%%%%%%%%%%%%%%%%%%%%%%%%%%%%%%%%%%%%%%%%%%%%%%%%%%%%%%%%%%%%%%%%%%%%%%

When $g_{3} g_{8} \not= 0$, it can be shown that $g(k, \eta) = 0$ can have at most three different turning points, 
depending on the signs of ${g}_{3}$, as shown in Fig. \ref{fig2}.  In particular, when ${g}_{3} \ge 0$, $g(k, \eta)$ has 
only one turning point. When  ${g}_{3} < 0$, $g(k, \eta)$ can have one, two or three turning points, depending on the 
ratio of ${g}_{3}/g_8$. In this paper, we consider only the cases where $g(k, \eta)$ has only one turing point. Then,
$g(k, \eta)$ given by Eq. (\ref{3.34}) can be written as
\bq
\lb{3.39}
g(k, \eta) = \frac{\tilde{g}_{8}k^{2}}{z^{2}}\Big(z^{2}_{0} - z^{2}\Big)\Big(z^{4} + az^{2} + b\Big),
\eq
where $z_{0} = |k \bar{\eta}(k)|$, and
\bq
\lb{3.40}
a \equiv \frac{\tilde{g}_{3}}{\tilde{g}_{8}} + z^{2}_{0},\;\;
b  \equiv \frac{1}{\tilde{g}_{8}} + a z^{2}_{0},\;\; bz_{0}^{2} = \frac{9}{4\tilde{g}_{8}}.
\eq
Since $g(k, \eta)$ has only one turning point, we must have $z^{4} + az^{2} + b > 0$.
 Inserting Eq. (\ref{3.39}) into Eq. (\ref{3.30a}), we find that
 \bq
 \lb{3.41}
 {\cal{D}}(k, \eta) = - \sqrt{\tilde{g}_{8}}\int^{z}_{z_{0}(k)}{\frac{\sqrt{z_{0}^{2} - z^{2}}}{z} G(z) dz},
 \eq
 where 
 \bq
 \lb{3.42}
 G(z) \equiv \sqrt{z^{4} + az^{2} + b} = \sqrt{b} + \frac{a}{2\sqrt{b}}z^{2} %\nb\\
% &=& \sqrt{b} + \frac{a}{2\sqrt{b}}z^{2} 
%& & + \frac{1}{2}\Bigg(\frac{1}{\sqrt{b}}
 %- \frac{a^{2}}{4b^{3/2}}\Bigg)z ^{4}  
 + {\cal{O}}\Big(z^{4}\Big).
 \eq
Substituting the above into Eq. (\ref{3.41}), we obtain
\bqn
\lb{3.43}
{\cal{D}}(k, \eta) &\simeq& - \frac{3}{2z_{0}}\sqrt{z_{0}^{2} - z^{2}}  - \frac{3}{2}\ln\left(\frac{\big(-z\big)}{z_0 +\sqrt{z_{0}^{2} - z^{2}} }\right)\nb\\
& & + \frac{z_0}{9}\Big(\tilde{g}_{3} + \tilde{g}_{8} z_{0}^{2}\Big)\Big(z_{0}^{2} - z^{2}\Big)^{3/2},
\eqn
from which we find that
\bq
\lb{3.44}
n_{T} \simeq \Bigg(1 + k \frac{\bar{\eta}'}{\bar{\eta}}\Bigg)\Bigg[3 + \frac{4k^{4}\bar{\eta}^{4}}{9}
\Bigg(\frac{2g_{3}H^{2}}{\zeta^{2}} + \frac{3|g_{8}|H^{4}}{\zeta^{4}}k^{2}\bar{\eta}^{2}\Bigg)\Bigg].
\eq
 When $g_3 = g_8 = 0$, we have $\bar{\eta} = - 3/(2k)$, and the above expression yields $n_T = 0$, which is exactly the result given by
 Eq. (\ref{3.38}). When $g_3 $ and $g_8$ are different from zero, the last two terms in the right-hand side represent corrections from
 the high order derivatives of the curvature, which are suppressed by the Planck scale $\zeta^{2} = M^{2}_{pl}/2$.   
 Inserting  Eq. (\ref{3.44}) into Eq.(\ref{3.29}), on the other hand, we find that 
 \bq
 \lb{3.45}
 P_{T}(k) = \frac{4H^{2}z^{3}_{0}}{3\pi^{2} e^{3}}\exp\Bigg\{ \frac{2H^{2}z^{4}_{0}}{9\zeta^{4}}
 \Big(g_{3}\zeta^{2} + |g_{8}|z^{2}_{0} H^{2}\Big)\Bigg\},
 \eq
 which is suppressed exponentially by the Planck scale.

 \subsection{The    Power-Law Background}
 
 When $a(t) \propto t^{1+n}$ or $a(\eta) = (-H\eta)^{-(1+1/n)}$, we find that
 \bq
 \lb{3.46}
 g(k, \eta) = k^{2}\Bigg\{\frac{\beta^{2}}{x^{2}} - \Big[1 + \hat{g}_{3} x^{2(1 + 1/n)} 
 +  \hat{g}_{8} x^{4(1 + 1/n)}\Big]\Bigg\},
 \eq
 where $x \equiv - H\eta$, and 
 \bq
 \lb{3.47}
 \beta \equiv \frac{(2+3n)H}{2nk}, \;\;\;  \hat{g}_{3} \equiv \frac{g_{3}k^{2}}{\zeta^{2}}, \;\;\; 
  \hat{g}_{8} \equiv \frac{|g_{8}| k^{4}}{\zeta^{4}}.
  \eq
Unlike that in GR, now inflation can be realized when $n > -2/3$ \cite{Muk}. The de Sitter 
universe corresponds to $n = \infty$.

When $g_{3} = g_{8} = 0$, we find that the zero of $g(k, \eta)$ is at $- k \bar{\eta} = \beta$, and Eq. (\ref{3.30a})
yields,
\bq
\lb{3.48}
{\cal{D}}(k, \eta) = - \frac{k}{H}\sqrt{\beta^{2} - x^{2}}
- \frac{3n+2}{2n}\ln\Bigg(\frac{x}{ \sqrt{\beta^{2} - x^{2}} + \beta}\Bigg).
\eq
Then, Eqs. (\ref{3.29}) and (\ref{3.30}) give
\bqn
\lb{3.49}
P_{T}(k) &=& \frac{1}{2\pi^{2} e^{3 + 2/n}} \left[\frac{(3n+2)H}{n}\right]^{2(1+1/n)} k^{-2/n},\nb\\
N_{T} &=& - \frac{2}{n}.
\eqn
As $n \rightarrow \infty$, the above expressions reduce to the ones given by Eq. (\ref{3.38}). 

When $g_{3}$ and $ g_{8}$ are different from zero and $\epsilon = k^{2}/\zeta^{2} \ll 1$, 
the zero of $g(k, \eta) = 0$ is well approximated by $x  = \beta$. Then, $g(k, \eta)$ can be written as
\cite{YKN},
\bq
\lb{3.50}
g(k, \eta) = \frac{k^{2}}{x^{2}}\Big(\beta^{2} - x^{2(1+ \nu)}\Big),
\eq
where 
\bqn
\lb{3.51}
\nu(k, x) &\equiv& \frac{1}{2}\frac{d\ln{\tilde{g}(x)}}{d\ln (x)} \nb\\
&=& \frac{1+n}{n\tilde{g}(x)}\Bigg(\epsilon g_{3}  + 2 |g_{8}| \epsilon^{2} x^{2(1+1/n)}\Bigg)x^{2(1+1/n)},\nb\\
\tilde{g}(x) &\equiv& 1 + \epsilon g_{3} x^{2(1+1/n)} +  |g_{8}| \epsilon^{2} x^{4(1+1/n)}.
\eqn
Thus, we find that
\bqn
\lb{3.52}
{\cal{D}}(k, \eta) &=& - \frac{k\beta}{(1+ \bar{\nu})H}
\Bigg[\frac{\sqrt{\beta^{2} - x^{2(1 + \bar{\nu})}}}{\beta} + \big(1+\bar{\nu}\big)\ln\frac{x}{\beta}\nb\\
& &
+ \ln\Bigg(\frac{\beta}{\beta + \sqrt{\beta^{2} - x^{2(1 + \bar{\nu})}}}\Bigg)\Bigg],
\eqn
where 
\bqn
\lb{3.52a}
\bar{\nu} &\equiv& \nu(k, \beta) = \frac{1+n}{n \zeta^{4}}\Bigg\{g_{3}\zeta^{2}k^{2}\beta^{2(1+1/n)}\nb\\
& & ~~~~~
+ \big(2|g_{8}|-g^{2}_{3}\big)k^{4}\beta^{4(1+1/n)}\Bigg\}.
\eqn
 From the above expressions we obtain
\bqn
\lb{3.53}
P_{T}(k) &\simeq& \frac{k^{-2/n}}{2\pi^{2}}\Bigg[\frac{(3n+2)H}{n}\Bigg]^{2(1 + 1/n)}\nb\\
& & ~~~~~~~~~~ \times 
\exp\Bigg\{- \frac{(3n+2)}{n(1+ \bar{\nu})}\Bigg\},\nb\\
n_{T} &\simeq&  - \frac{2}{n} - \frac{2(1+n)(2+3n)(1-\ln 2)}{n^{3}(1+ \bar{\nu})^{2}\zeta^{2}k^{4/n}}  \Bigg\{g_{3} k^{2/n} \nb\\
& & + \frac{2\big(2|g_{8}| - g^{2}_{3}\big)}{\zeta^{2}}
\Bigg[\frac{(3n+2)H}{2n}\Bigg]^{2(1 + 1/n)}\Bigg\}\nb\\
& & \times \Bigg[\frac{(3n+2)H}{2n}\Bigg]^{2(1 + 1/n)}. 
\eqn

When   $\epsilon = k^{2}/\zeta^{2} \gg 1$, on the other hand, 
the zero of $g(k, \eta) = 0$ is well approximated by 
\bq
\lb{3.54}
\bar{x} = - H \bar{\eta} = \left[\frac{(3n+2)H\zeta^{2}}{2nk^{3}\sqrt{|g_{8}|}}\right]^{\frac{n}{3n + 2}}.
\eq
Then, $g(k, \eta)$ can be written as
\bq
\lb{3.55}
g(k, \eta) = \frac{k^{6}}{x^{2}\zeta^{4}}\Big(\tilde{\beta}^{2} - x^{2(1+ \nu)}\Big),
\eq
but now with
\bqn
\lb{3.56}
\tilde{\beta} &=& \frac{(3n+2)\zeta^{2}H}{2nk^{3}},\nb\\
\nu(k, x) %&\equiv& \frac{1}{2}\frac{d\ln{\tilde{g}(x)}}{d\ln (x)} \nb\\
&=& \frac{1+n}{n\tilde{g}(x)}\Bigg(2 |g_{8}| x^{2(1+1/n)} + \frac{g_{3}}{\epsilon} \Bigg)x^{2(1+1/n)},\nb\\
\tilde{g}(x) &\equiv&   |g_{8}|   x^{4(1+1/n)}  +  \frac{g_{3}}{\epsilon} x^{2(1+1/n)} +  \frac{1}{\epsilon^{2}}.
\eqn
Then, we find that
\bqn
\lb{3.57}
{\cal{D}}(k, \eta) &=& - \frac{3n + 2}{2n(1+ \bar{\nu})}
\Bigg[\frac{\sqrt{\tilde{\beta}^{2} - x^{2(1 + \bar{\nu})}}}{\tilde{\beta}} + \big(1+\bar{\nu}\big)\ln\frac{x}{\tilde{\beta}}\nb\\
& &
+ \ln\Bigg(\frac{\tilde{\beta}}{\tilde{\beta} + \sqrt{\tilde{\beta}^{2} - x^{2(1 + \bar{\nu})}}}\Bigg)\Bigg],
\eqn
where $\bar{\nu} \equiv \nu(k, \bar{x})$, from which we obtain
\bqn
\lb{3.58}
P_{T}(k) &\simeq& \frac{\zeta^{2}}{2\pi^{2}}\Bigg[\frac{(3n+2)\zeta^{2}H}{nk^{3}}\Bigg]^{2(1 + 1/n)}\nb\\
& & ~~~~~~~~~~ \times 
\exp\Bigg\{- \frac{(3n+2)}{n(1+ \bar{\nu})}\Bigg\},\nb\\
n_{T} &\simeq&  - \frac{6(1+n)}{n} %- \frac{4(1-\ln 2)(1+n)\zeta^{2}}{n^{2}(1+ \bar{\nu})^{2}g_{8}^{2}} 
 \nb\\
& &  + \frac{4(1-\ln 2)(1+n)\zeta^{2}}{n^{2}(1+ \bar{\nu})^{2}g_{8}^{2}}  
\Bigg[\frac{2n\sqrt{|g_{8}|}}{(3n+2)\zeta^{2}H}\Bigg]^{\frac{2(1 + n)}{3n+2}}\nb\\
& & \times \Bigg\{\zeta^{2}\big(g^{2}_{3} - 2 |g_{8}|\big) \Bigg[\frac{2n\sqrt{|g_{8}|}}{(3n+2)\zeta^{2}H}\Bigg]^{\frac{2(1 + n)}{3n+2}}\nb\\
& & ~~~~~~~~ - g_{3} |g_{8}| \Bigg\} k^{\frac{4}{3n + 2}}.
\eqn

\section{Conclusions}

We have studied  cosmological vector and tensor perturbations in the most general SVW setup of the HL theory with the projectability
condition but without the detailed balance. For the vector perturbations, we have showed explicitly that the resulted expressions are
identical to those given in GR. Thus, all the results obtained in GR regarding to the vector perturbations also hold here in the HL
theory. 

For the tensor perturbations, we found that, among other things, the high order derivatives of curvatures produces an effective stress,
which could produce high-frequency gravitational waves even when the matter anisotropic stress vanishes. These terms have negligible
efforts on the low-frequency modes of gravitational waves. The dispersion relations contain three different terms, proportional to, respectively,
$k^{2},\; k^{4}$ and $k^{6}$. As a result, in the UV regime the power spectrum is scalar-invariant, while in the IR limit it eventually reduces 
to that given in GR. 

Applying our general formulas for tensor perturbations to the background of de Sitter as well as the power-law expansions, we were
able to calculate the corresponding power spectra and indices analytically, using the uniform approximations proposed recently
by  Habib {\em et al} \cite{Habib}.

In this paper, we  assumed that the strong-coupling problem \cite{CNPS,KA,PS} in cosmological backgrounds can also be addressed 
via the Vainshtein mechanism \cite{Vain,VainPRD} or some other approach. We wish to come back to this issue soon.

\begin{acknowledgments}

  The author would like to express his gratitude to Roy Maartens and David Wands for valuable discussions and suggestions, and their
  collaboration in the early stage of this work, which was supported in part by  DOE Grant, DE-FG02-10ER41692.
  
\end{acknowledgments}

%%%%%%%%%%%%%%%%%%%%%%%%%%%%%%%%%%%%%%%%%%%%%%%%%%%%%%

\end{document}